\documentclass[useAMS,useusegraphicx]{mn2e}
\usepackage{amssymb}
\usepackage{epsfig}
\def\etal{et al.\ }
\def\aj{AJ}

\def\apj{ApJ}
\def\apjl{ApJ}
\def\aap{A\&A}
\def\mnras{MNRAS}

\def\pasj{PASJ} 

\usepackage{epsfig}
\newif\ifAMStwofonts

\title[Bulges and Bars in S0 galaxies]
{Which bulges are favoured by barred S0 galaxies?}
\author[Barway \etal]
{Sudhanshu Barway $^{1}$\thanks{E-mail: barway@saao.ac.za (SB)}, Kanak
  Saha $^{2}$\thanks{E-mail: kanak@iucaa.in (KS)}, Kaustubh Vaghmare
  $^{2}$\thanks{E-mail: kaustubh@iucaa.ernetin (KV)}, 
\newauthor
 and  Ajit K. Kembhavi $^{3}$\thanks{E-mail: akk@iucaa.ernet.in (AKK)}\\
$^{1}$South African Astronomical Observatory, P.O. Box 9, 7935, Observatory, Cape Town, South Africa; \\ 
$^{2}$Inter University Centre for Astronomy and Astrophysics, Post Bag 4, Ganeshkhind, Pune 411007, India. \\  } 

\begin{document}
\maketitle
\label{firstpage}

\begin{abstract}
S0 galaxies are known to host classical bulges with a broad range of size and
mass, while some such S0s are barred and some not.   The origin of the
bars has remained as a long-standing problem -- what made bar formation possible in certain S0s?

By analysing a large sample of S0s with classical bulges observed by the Spitzer space telescope,
we find that most of our barred S0s host comparatively low-mass classical bulges,
typically with bulge-to-total ratio ($B/T$) less than $0.5$; whereas S0s with
more massive classical bulges than these do not host any bar. Furthermore, we
find that amongst the barred S0s, there is a trend for the longer and massive
bars to be associated with comparatively bigger and massive classical
bulges -- possibly suggesting bar growth being facilitated by these classical bulges.
In addition, we find that the bulge effective radius is always less than the
bar effective radius --indicating an interesting synergy between the host
classical bulge and bars being maintained while bar growth occurred in these S0s.
\end{abstract}


\begin{keywords}

galaxies: elliptical and lenticular -   fundamental parameters
galaxies: photometry - structure - bulges 
galaxies: formation - evolution
\end{keywords}


\section{Introduction}
\label{sec:intro}

Lenticular (S0) galaxies in the local universe are primarily
characterised by the presence of a bulge and disc with no apparent  
spiral arms \nocite{Barwayetal2007,Vandenbergh2009}({Barway} {et~al.} 2007; {van den Bergh} 2009) - but a number of 
observations have shown that like their progenitor spirals, S0 galaxies, especially the low
luminous ones, are both barred and unbarred \nocite{Barwayetal2011,vandenBergh2012}({Barway}, {Wadadekar} \&  {Kembhavi} 2011; {van den Bergh} 2012). 
What has made bar formation possible in some S0 galaxies has remained
a long standing puzzle.

Significant progress has been made over the last decade or so in terms of our 
understanding of the redshift evolution of bars in disc galaxies. A number of these
studies suggest that the bar fraction in spiral galaxies is strongly dependent on their mass
\nocite{NairAbraham2010, Cameronetal2010}({Nair} \& {Abraham} 2010; {Cameron} {et~al.} 2010). It has been shown that the bar fraction in
low-mass spirals remains nearly constant out to $z \sim 1$, corresponding to a look-back time of 
$7.8$~billion years \nocite{Elmegreenetal2004,Jogeeetal2004,Barazzaetal2008, 
NairAbraham2010, Cameronetal2010}({Elmegreen}, {Elmegreen} \&  {Hirst} 2004; {Jogee} {et~al.} 2004; {Barazza}, {Jogee}, \&  {Marinova} 2008; {Nair} \& {Abraham} 2010; {Cameron} {et~al.} 2010). More recently, \nocite{Simmonsetal2014}{Simmons} {et~al.} (2014) using the 
HST CANDELS survey extended such a study to $z \sim 2$ and found no significant change in the 
bar fraction. These findings imply that bars are robust stellar structures; 
once formed, it is hard to destroy them. Based on the modelling of stellar kinematics, it is 
believed that the barred spirals were the progenitors of the present-day barred lenticulars
 which got rid of their spirals \nocite{Cortesietal2011, Cortesietal2013}({Cortesi} {et~al.} 2011, 2013) - it becomes clearer 
that bars in the present-day S0s have formed long back, most likely during the cosmic assembly
of disc galaxies. During those early phase of evolution, a disc would have assembled and grown 
around a classical bulge either merger-built \nocite{Kauffmanetal1993, Baughetal1996, 
Hopkinsetal2009}({Kauffmann}, {White} \&  {Guiderdoni} 1993; {Baugh}, {Cole}, \& {Frenk} 1996; {Hopkins} {et~al.} 2009) or formed as a result of other processes likely to be active in the high-redshift 
universe e.g., clump coalescence, violent disc instability etc. 
\nocite{Elmegreenetal2008,Ceverinoetal2015}({Elmegreen}, {Bournaud} \&  {Elmegreen} 2008; {Ceverino} {et~al.} 2015). Then one would expect the classical bulge to 
intervene the bar formation process that occurred in the host stellar disc of the present-day S0s. 

Indeed, \nocite{Barazzaetal2008}{Barazza} {et~al.} (2008) showed that bar fraction rises sharply from 
$\sim 40 \%$ to $70\%$ as one moves from early-type to late-type
galaxies which are disc dominated rather than ones with 
prominent bulges. A massive classical bulge can produce a
strong inner Lindblad resonance (ILR) barrier to
prevent the feedback loop required for the swing amplification mechanism to work
effectively in the disc leading to the formation of a bar in the first 
place \nocite{Dubinskietal2009}({Dubinski}, {Berentzen} \&  {Shlosman} 2009).
So it is desirable for a stellar disc to not have a strong ILR in the early phase 
of galaxy assembly. A massive classical bulge can also produce enough central 
concentration to create destructive effect on the orbital backbones of a bar 
\nocite{PfennigerNorman1990, Hasanetal1993}({Pfenniger} \& {Norman} 1990; {Hasan}, {Pfenniger} \&  {Norman} 1993). Overall, it turns out that a massive 
classical bulge and a bar might not coexist in a spiral galaxy. But it remains
unclear how to reconcile this with the observed properties of bars and classical
bulges in S0 galaxies. The primary aim of the current work is to understand
what physical parameters of a classical bulge are a pre-requisite for a bar to form and
grow stronger in a S0 galaxy. 

The paper is organised as follows. Section~\ref{sec:data} describes the sample data
and its analysis. The role of S0 discs and classical bulges 
in the context of bar formation are considered in section~\ref{sec:disc} and
section~\ref{sec:ClB}. Section~\ref{sec:discuss} is devoted to discussion and conclusions.  
Throughout this paper, we use the standard concordance cosmology with $\Omega_M= 0.3$, 
$\Omega_\Lambda= 0.7$ and $h_{100}=0.7.$

\begin{figure}
\rotatebox{0}{\includegraphics[height=7.5cm]{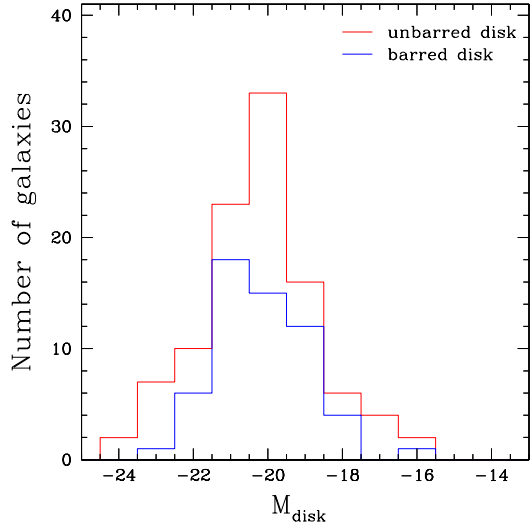}}
\caption{Distribution of absolute magnitudes (in 3.6 $\mu m$) of the host stellar discs in barred and 
unbarred S0 galaxies. Both barred and unbarred discs seem to have similar range of 
disc luminosities.}
\label{fig:Mdisc}
\end{figure}

\section{Sample}
\label{sec:data}

The sample used in this Letter is described in detail in \nocite{vaghmare2015}{Vaghmare} {et~al.} (2015).
The parent sample comprises 1031 galaxies, visually classified as S0 in the RC3 catalogue
\nocite{rc1991}({de Vaucouleurs} {et~al.} 1991) and having an integrated B-band magnitude brighter than
14.0. This parent sample is cross-matched with the Spitzer Heritage Archive
(SHA) and $247$ galaxies are found with 3.6 $\mu m$ imaging data. The Level 1 or
Basic Calibrated Data were obtained from SHA and a co-added mosaic was
constructed using MOPEX (MOsaicking and Point EXtraction tool). Structural
parameters for the bulge, disc and bar components of these galaxies were derived
using GALFIT \nocite{peng2002}({Peng} {et~al.} 2002) on the mosaics. In
the first run, we fitted all galaxies with a bulge and a disk using a S\'{e}rsic 
\nocite{Sersic68}({Sersic} 1968) and an exponential profile simultaneously.
In cases, where the residual image  obtained by subtracting the Point
Spread Function convolved best-fit model from the observed image revealed a
bar, a second run of fitting was performed by adding another Sersic
component to describe the bar. Inclusion of an additional bar component
do not show any difference for  S\'{e}rsic index $n$ values  for both barred and
unbarred classical bulges,  respectively.  We also found some galaxies
that were visually classified as S0s, had spiral like features in GALFIT residual images. We have removed them, 
in addition to those with bad fits, poor quality images from our subsequent analysis.
Our final sample comprises of $185$ S0 galaxies with  median redshift
of  0.005.

In order to classify the bulges, the authors, in  \nocite{vaghmare2013}{Vaghmare}, {Barway} \&  {Kembhavi} (2013), used a
combination of two well established criteria in the literature. All bulges
deviating more than $3-\sigma$ below the best-fit line to the Kormendy relation to
ellipticals \nocite{Gadotti2009}({Gadotti} 2009) and having a S\'{e}rsic index $n < 2$ \nocite{FisherDrory2008}({Fisher} \& {Drory} 2008) 
were classified as pseudo bulges, while the rest were classified as classical bulges. The sample
comprises 25 pseudo bulge hosts with the remaining the 160 being classical bulge
hosts. 65 of 160 classical bulge hosts are barred. To determine 
the masses of the bulges and discs in our galaxies, we used a
prescription by \nocite{cook2014}{Cook} {et~al.} (2014) to obtain the $M/L$ ratios at 3.6 $\mu
m$. In this paper, we consider the relationship between classical
bulges and bars. Our current sample does not have a large enough number of pseudo bulges to investigate
their relationship to bars in S0 galaxies (see
\nocite{vaghmare2015}{Vaghmare} {et~al.} (2015)). Hereafter, we refer a classical bulge as a bulge
unless stated otherwise. 

\begin{figure}
\rotatebox{0}{\includegraphics[height=7cm]{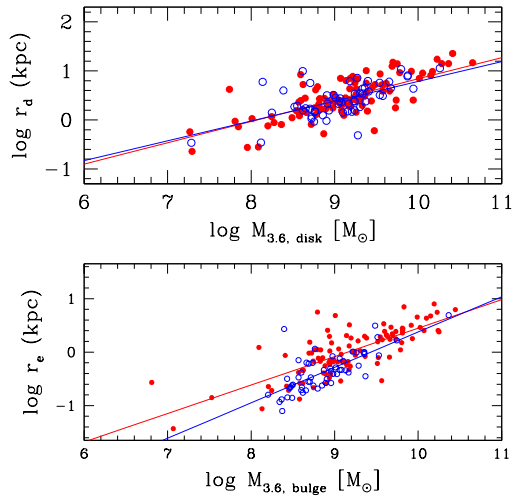}}
\caption{Top panel: Size -mass relation for the disc hosting classical bulges in the 3.6 $\mu m$.
Filled red circles are  unbarred S0s and open blue circles
are barred S0s. Solid lines are best fit line to the data. 
Bottom panel: the same for the classical bulges only (unbarred and barred).}
\label{fig:size-mass}
\end{figure}
\begin{figure*}
\rotatebox{0}{\includegraphics[height=6cm]{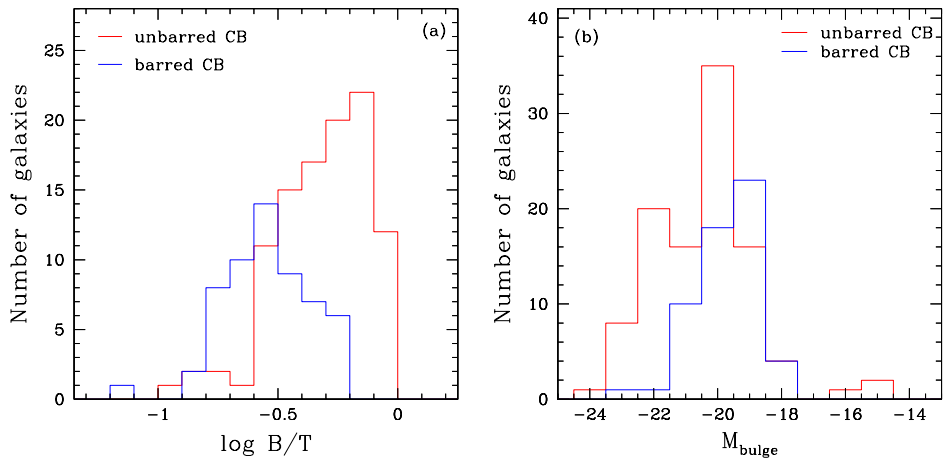}}
\caption{{\bf a.} Bulge-to-total ratios ($B/T$) distribution for classical bulges. Bars are associated with comparatively low $B/T$s.
{\bf b.} Absolute magnitude distribution for the classical bulges. }
\label{fig:BbyT}
\end{figure*}

\section{Role of S0 discs on bar formation}
\label{sec:disc}
It is well known that massive cool self-gravitating discs are, in general, 
prone to bar instability - leading to the formation of strong bars as shown 
by a number of simulations \nocite{Hohl1971,DebattistaSellwood1998, Athanassoula2003, 
Sahaetal2012}({Hohl} 1971; {Debattista} \& {Sellwood} 1998; {Athanassoula} 2003; {Saha}, {Martinez-Valpuesta} \&  {Gerhard} 2012). Whereas lower mass, comparatively hotter galaxies tend to avoid forming
strong bars \nocite{Sahaetal2010, Shethetal2012, Saha2014}({Saha}, {Tseng} \& {Taam} 2010; {Sheth} {et~al.} 2012; {Saha} 2014). When it comes to S0 galaxies 
in the local universe, bars are preferentially observed in low-luminous S0s 
\nocite{Barwayetal2011}({Barway} {et~al.} 2011) - which is a puzzling issue. Here we study the global properties of 
stellar discs of S0s to disentangle their role played in the formation of a bar.   

A three-component (bulge-bar-disc) decomposition of each S0 galaxy provides us the structural 
information about stellar discs. All stellar discs are well modelled by an exponentially falling
surface brightness distribution with scale length $r_d$ and central surface brightness $I_0$.
Using these and the distance information, we compute the absolute magnitude $M_{disc}$ of the
disc in the $3.6$ $\mu m$. In Fig.~\ref{fig:Mdisc}, we show the distribution of 
$M_{disc}$ for $160$ galaxies with 
classical bulges. {\it This sample has been sub-divided into two  - a
  subsample of barred galaxies and that of unbarred.} What we notice is that, overall, the histograms of disc 
luminosity appear remarkably similar in either case. In other words, we do not find 
any particular luminosity range being preferred by an S0 disc to host a bar.
 This is also being reflected in the top panel of Fig.~\ref{fig:size-mass}, showing the relation 
between the disc scale length and disc stellar mass - the so called
{\it size-mass relation} \nocite{Gadotti2009}({Gadotti} 2009). {\it We find that both barred and unbarred 
stellar discs follow nearly the same size-mass relation.} In other words, where bar formation 
is concerned, there is no circumstantial preference for either low-mass (hence low luminous) or 
high-mass discs in the current S0 sample. However, a conclusive remark on this aspect requires
one to probe an even larger sample of stellar discs without an
environmental or morphological bias. In the following, we investigate
whether the incidence of a bar  in our sample S0s depends on the 
presence of a classical bulge.    

\section{Classical bulge and bar connection}
\label{sec:ClB}
Classical bulges and bars coexist in spiral galaxies across the Hubble sequence from
late-type to early-type and S0s. Yet, it remains to be established whether
classical bulges play any role in the bar formation. Part of the difficulty lies in
disentangling whether bars formed after the classical bulges, or both formed nearly
simultaneously, or bars existed before the classical bulges formed. We consider the first
as a viable scenario - as might have been the case if major mergers formed a classical bulge
and the disc assembled around it gradually and became self-gravitating leading to the formation
of bars. All the S0s in the current sample that host classical bulges, the bar being the
only morphological entity used to classify them into two categories - barred and unbarred.
In other words, the S0 discs plus classical bulge acts as a base
structure. In some cases, this structure allows for a bar to grow and
in some cases, it does not. On what aspects of this base structure
does the bar formation depend on, remains one of the outstanding
issues in astronomy. Below, we attempt to unravel a link between
classical bulge properties and growth of bar in the stellar disc.

First, we study the bulge-to-total ratios $B/T$ in our current
sample (see Fig.~\ref{fig:BbyT}(a)).
We find that the distribution of $B/T$ is clearly separated for barred and unbarred 
S0s, with mean $B/T \sim 0.35$ for barred S0s and $ \sim 0.7$ for unbarred ones.
Although $\sim 30\%$ of the total luminosity in the bulge is substantial compared to late-type
galaxies, where S0s are concerned, this is considered to be small. So we calculated
the absolute bulge luminosity and found that the bulges in barred S0s are actually 
of lower luminosity compared to those in unbarred ones by $\sim 1$ magnitude (Fig.~\ref{fig:BbyT}(b)).
Bottom panel of fig.~\ref{fig:size-mass} shows the size-mass relation for all the 
classical bulges. It is obvious from the figure that both the
effective radii $r_e$ and stellar mass of classical bulges are smaller in S0s
with bars than those without any bar. A large fraction of the classical bulges in barred
S0s have masses falling in the range $\sim 10^{8} - 10^{9} M_{\odot}$ while 
their disc mass scatter around few~$\times 10^9 M_{\odot}$.
In other words, {\it we do see a preference of smaller
low-mass classical bulges to be associated with barred S0s over unbarred ones.}
This, in turn, would imply that perhaps bar formation was hindered in 
galaxies with massive bulges in the central region. Since a massive bulge would
produce a strong inner Lindblad resonance (ILR) which might cut the feedback loop
necessary for the swing-amplification  \nocite{Toomre1981}({Toomre} 1981) to work efficiently and thereby slowed down or even 
stopped growing a bar in the first place. 
\begin{figure}
\rotatebox{0}{\includegraphics[width=7cm]{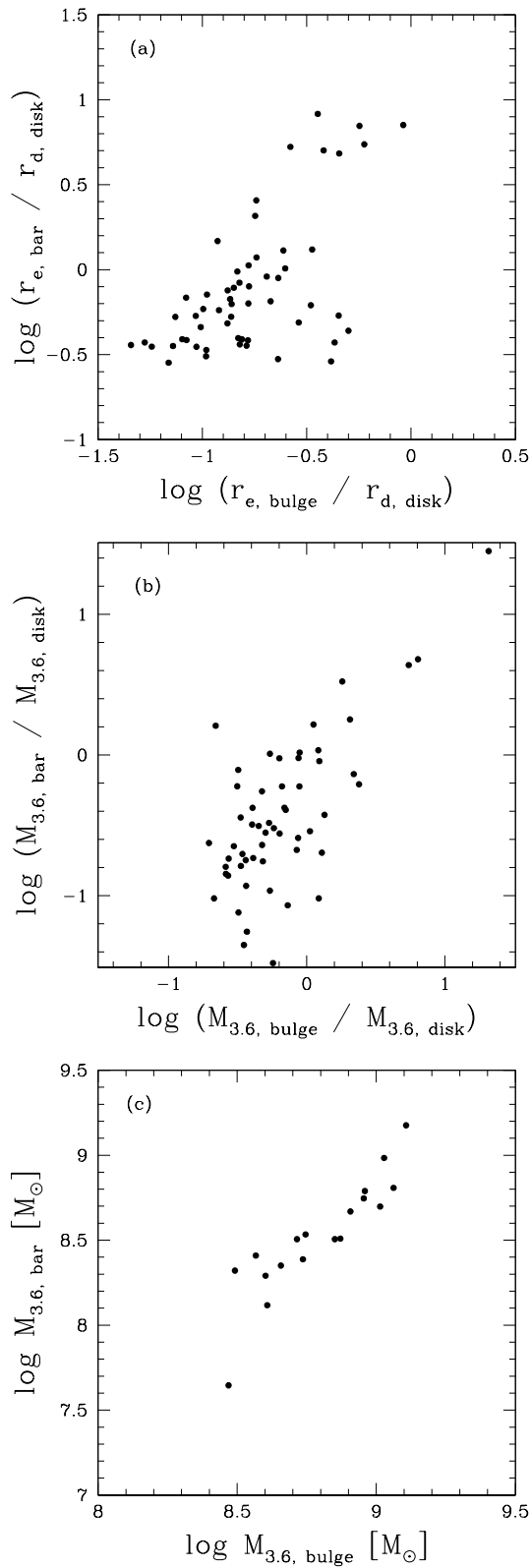}}
\caption{Correlation between the bar and classical bulge parameters for barred S0 galaxies.
{\bf a.} Correlation between bar effective radii $r_{e, Bar}$ and bulge effective
radii $r_{e, bulge}$, normalised by the disc scale lengths $r_d$.
{\bf b.} Normalised bar mass plotted against normalised bulge mass. {\bf c}. Bulge mass
versus bar mass in units of $M_{\odot}$ for those S0s with nearly same 
disc mass $\sim$ $2 \times 10^{9} M_{\odot}$.}
\label{fig:barbulge} 
\end{figure}
\subsection{Correlation between bar-bulge parameters}
\label{sec:corr}
The idea here is to find out whether there is a deeper connection between a bar
and a classical bulge in a barred S0 galaxy. Fig.~\ref{fig:barbulge}~(a) shows that there is
a trend for longer bars to be associated with bigger bulges. It is interesting to note that 
effective radii of the classical bulges are less than or equal to the disc scale length 
whereas bars come with a wider range of sizes. In addition to the bar-bulge size correlation,
 Fig.~\ref{fig:barbulge}~(b) reveals a clear trend of massive bars being associated with 
massive classical bulges amongst the barred S0s. Taken together, it implies that
longer and massive bars are developed in those S0s which host bigger and comparatively 
massive classical bulges. In other words, {\it massive, bigger classical bulges seem to facilitate
bars to grow longer and be more massive.} Note that this holds true only for the barred S0s having
$B/T < 0.5$ as found in the current sample. To strengthen this further, {\it we constructed a 
subset of barred S0s whose stellar disc masses were nearly equal. Fig.~\ref{fig:barbulge}~(c) 
shows a rather strong correlation between the bulge mass and bar mass for those S0s.} What
might have happened is that these comparatively low $B/T$ classical bulges allowed the
bar formation in the first place because of lowered ILR strength and subsequently facilitated
the bar growth via resonant gravitational interaction \nocite{Sahaetal2012, sahaGerhard2013, 
Sahaetal2015}({Saha} {et~al.} 2012; {Saha} \& {Gerhard} 2013; {Saha}, {Gerhard} \&  {Martinez-Valpuesta} 2016). Needless to say the surrounding dark matter halo would
also play a similar role alongside but we do not have any information about the halo at 
this point, except input from simulations, see
section~\ref{sec:discuss} below. 

\section{Discussion}
\label{sec:discuss}

The role of a classical bulge in the growth and evolution of a bar has not been
fully investigated. It is known that a massive centrally concentrated object (e.g.,
a supermassive black hole) can considerably weaken a bar by scattering stars off the
$x_1$-family of orbits which constitute the backbone of the bar \nocite{Hasanetal1993,
SellwoodMoore1999, Athanassoulaetal2005,HozumiHerquist2005,Hozumi2012}({Hasan} {et~al.} 1993; {Sellwood} \& {Moore} 1999; {Athanassoula}, {Lambert} \&  {Dehnen} 2005; {Hozumi} \& {Hernquist} 2005; {Hozumi} 2012).
Some of the classical bulges in the current sample have the right mass
for such action, but are not as centrally concentrated as might be
required to have a supermassive black hole like  effect . But such
massive classical bulges could, in principle, delay or even stop a bar from forming 
in the first place by producing an ILR near the centre of the galaxy,
which would cut the feedback loop required for the swing amplification \nocite{Toomre1981}({Toomre} 1981). 
In fact, as mentioned in section~\ref{sec:ClB}, we do find S0 galaxies with massive 
classical bulges as unbarred. This agrees with  the reported
\nocite{Barazzaetal2008}{Barazza} {et~al.} (2008)  bar fraction reduction in disc galaxies with rising bulge-to-disc mass ratio.

It remains unclear why some spiral galaxies are barred and some are  not. Not only
spiral galaxies, but as we see here, S0s also face the same unresolved
issue. In order to make progress, one has to disentangle the effect of various parameters
of the disc, classical bulge and dark matter halo which determine the bar growth in a galaxy.  
N-body simulations have shown that a bar forms and grows rapidly in a cool, rotating 
self-gravitating disc \nocite{Hohl1971,SellwoodWilkinson1993,Athanassoula2002,
Dubinskietal2009, Sahaetal2012}({Hohl} 1971; {Sellwood} \& {Wilkinson} 1993; {Athanassoula} 2002; {Dubinski} {et~al.} 2009; {Saha} {et~al.} 2012, and references therein). Furthermore, the bar continues to grow in size and mass by 
transferring angular momentum from the inner disc to the surrounding dark matter halo via resonant 
gravitational interaction \nocite{DebattistaSellwood1998,Athanassoula2002,Holley-Bockelmannetal2005,
WeinbergKatz2007a,Ceverinoklypin2007,SahaNaab2013}({Debattista} \& {Sellwood} 1998; {Athanassoula} 2002; {Holley-Bockelmann}, {Weinberg} \&  {Katz} 2005; {Weinberg} \& {Katz} 2007; {Ceverino} \& {Klypin} 2007; {Saha} \& {Naab} 2013). But if the initial
disc was hotter and dark matter dominated, a bar would grow rather
slowly over several billion years and might remain weak and be  too
faint to be detected \nocite{Sahaetal2010, Shethetal2012,
  Saha2014}({Saha} {et~al.} 2010; {Sheth} {et~al.} 2012; {Saha} 2014). These two inputs lead us to suggest that the bars in S0s are unlikely to have formed in
their later phase of evolution. In other words, we think that bars in S0s formed during
the early phase of disc assembly around a classical bulge with a
comparatively lower $B/T$. Bars in S0 galaxies are preferentially formed in the presence of classical
bulges with lower $B/T < 0.5$. These classical bulges have their stellar mass in the range
 $\sim 10^8 - 10^9 M_{\odot}$. Massive classical bulges with $B/T > 0.5$ are not found in any barred
S0s in our sample. Amongst barred S0s with similar disc mass, there exist a strong correlation between
the bar and classical bulge properties. The host stellar discs are unlikely to have played a major role in the formation
of bars in these S0s.


\section*{acknowledgements}

We thank the anonymous referee for insightful comments that have
improved both the content and presentation of this paper. SB would 
like to acknowledge support from the National Research Foundation 
research grant (PID-93727). SB and KS acknowledge support from a 
bilateral grant under the Indo-South Africa Science and Technology 
Cooperation (PID-102296) funded by Departments of Science and 
Technology (DST) of the Indian and South African Governments




\end{document}